\documentclass[letterpaper]{article}
\usepackage{amsmath,amssymb,graphicx,enumerate,footnote}
\usepackage{verbatim,graphicx}
\usepackage{graphicx,color,verbatim,enumerate}
\usepackage{algorithmic,algorithm}
\begin{document}
\title{Design and Development of a User Specific Dynamic E-Magazine}
\author{Vikram Santhalia and Sanjay Singh\thanks{Sanjay Singh is with the Department of Information and Communication Technology, Manipal Institute of Technology, Manipal University, Manipal-576104, INDIA, E-mail: sanjay.singh@manipal.edu}}
\maketitle
\begin{abstract}
Internet and electronic media gaining more popularity due to ease and speed, the count of Internet users has increased tremendously. The world is moving faster each day with several events taking place at once and the Internet is flooded with information in every field. There are categories of information ranging from most relevant to user,  to the information totally irrelevant or less relevant to specific users. In such a scenario getting the information which is most relevant to the user is indispensable to save time.
\par
The motivation of our solution is based on the idea of optimizing the search for information automatically. This information is delivered to user in the form of an interactive GUI. The optimization of the contents or information served to him is based on his social networking profiles and on his reading habits on the proposed solution. The aim is to get the user's profile information based on his social networking profile considering that almost every Internet user has one. This helps us personalize the contents delivered to the user in order to produce what is most relevant to him, in the form of a personalized e-magazine. Further the proposed solution learns user's reading habits for example the news he saves or clicks the most and makes a decision to provide him with the best contents.
\end{abstract}


\section{Introduction}

Computers and Internet becoming an integral part of our lives, we have access to an abundance amount of information, anytime and anywhere making electronic media an important aspect. Within seconds and minutes of an event's occurrence it is recorded on the Internet giving us access to fresh and abundant data. Due to such a huge collection of information, it is time consuming to search for news or events we are really interested in knowing. We must also understand that everybody has different preferences and would have interest in different fields or topics. For instance, we can observe that when anything of major political importance crops up or an international sports team wins a big event, our social networking profiles are flooded with status updates, posts, links, comments etc. But the importance and interest for these feeds differs from user to user. Our aim here is to serve the user with the most relevant information using the proposed solution, the ''E-magazine''. The information served to the user is not restricted just to news feeds, the information ranges from news feeds to articles to blogs to movie review, best and latest apps from Android and Apple appstore to the alerts of upcoming sporting events etc.
\par
 Social media is gaining much popularity these days and almost every Internet user has a social networking account like Facebook, LinkedIn, Twitter, Google+ etc. These accounts hold a lot of personal information about the user that the user uploads himself making this information authentic and most updated. This profile information if analyzed could help understand the personal and professional life of the user. We can understand his likes, dislikes, recreational and professional interests etc. too.
\par 
In a competitive world like today's, optimal use of one's time is very important. Therefore, we need a product that can help to reduce the time of search for useful information.
\par
Rise of Internet led to the problem of information explosion \cite{elec13}. This is a state of rapid increase in published information available on the web in various media formats. Information explosion led to information pollution. Information pollution is the contamination of information supply with irrelevant, redundant, unsolicited and low-value information. The spread of useless and undesirable information has been demonstrated to impair decision making processes.
The result of this information explosion and pollution has been information overload, a condition of our time, where a person has difficulty understanding an issue and making decisions because they are overwhelmed with the presence of too much information.
It is like sensory overload for the information age. It is the access to too much information, almost instantaneously, without knowing the validity or quality of that information. This condition can lead to decision paralysis, where the person is unable to make a judgment as they cannot see what is relevant anymore.
\par 
Our proposed and developed software product ''User specific dynamic e-magazine'' is the end to these problems. It cuts down the explosion by giving the best contents. Filtering, clustering and classification of contents removes the pollution in the information. Getting the contents from hundreds of trusted sources and displaying only the concise form of the information brings an end to information overload.
\par
The proposed solution is an web-application where the users have to sign-up or register with some of their basic information like email-id etc. This web-application runs on all modern browsers available. It links user's social networking sites after gaining user's permission. Once the web-application is given permission by the user to connect with his social networking profiles, user's data from his profiles are picked and analyzed. His interests are calculated applying various algorithms on the data collected from his social networking profiles. After which the user is served with the feeds and news of his interests. The user can also view and customize his interests for a better experience. Once the feeds are generated, they are shown to the user in a very interactive GUI, which is in the form of a personal e-magazine. The feeds that come up on the user's magazine are dynamic and gets updated automatically. User still has an option to save the feeds which he does not wants to lose or may read later. He can directly search for some feeds, which is an in built feature of this application. The feeds which appear on his profile can be shared on his social networking sites just by the click of a button, he can even mail the feeds to someone or himself. Again this behavior of the user to read his feeds are analyzed i.e. the application analyzes the feeds that he shares on his social networking profiles, saves, mails to someone, clicks on the links to read more about a particular field, or searches for some feeds. This further enhances the accuracy and dynamicity of the application to serve the best contents for the user.

\par
The rest of this paper is organized as follows. Section \ref{a} gives the brief details of existing products similar to the proposed solution. Section \ref{b} explains the development of the proposed application E-magazine. Finally section \ref{c} concludes the paper.

\section{Related Work}
\label{a}

There are a few applications existing which are based on user interests. In this section we will discuss how the proposed solution is technically advanced and better than the ones available.

\par
StumbleUpon \cite{elec15} is a discovery engine (a form of web search engine) that finds and recommends web content to its users. Its features allow users to discover and rate Web pages, photos, and videos that are personalized to their tastes and interests using peer-sourcing and social-networking principles.
StumbleUpon produces decent search results, but the problem is with the dynamicity. Users have to manually select or choose their interests and web-contents are then selected and produced on the basis of the interests provided by the users. So, defining interests each time is next to searching feeds or contents on the web manually. The punch of StumbleUpon is that user's can provide their interests and StumbleUpon will generate the best contents according to the interest keywords provided by the user. The contents are fairly filtered but the dynamicity is lacking, giving the interests manually each time looks like a cumbersome task.

\par
Enliten \cite{elec16} collects news and information from many sources and delivers it in the form of a personal newspaper with engaging content. It has an option to plugin the social media accounts, and aggregate information streams according to the interests given by the user. It has an attractive GUI in the form of a newspaper and it learns user interest from ratings given by the user and from the information taken from his social networking sites.
\par
Moreover it does not have the features to save a news feed to read in the future or mail it to a friend. The problem with Enliten is that it produces only news feeds, whereas there is a scope to filter various other information available on the web apart from news feeds. 
\par
Zite \cite{elec17} evaluates new stories every day, looking at the type of article, its key attributes and how it is shared across the web. Zite uses this information to match stories to user's personal interests and then delivers them automatically to their iPad or iPhone. It learns by tapping into user's Google reader and twitter account history. It learns user's reading habits from the way user interacts with the articles he reads. However, it lacks several features like searching, saving, sharing etc and most importantly an attractive GUI.
\par
Last but not the least, we will talk about the social networking giant Facebook's new feature Interest Lists \cite{elec18}. Interest lists are an optional way to organize the content a particular user is interested in on Facebook. The user can create their own interest lists based on the things they care about, or subscribe to other people's lists. When the user creates or subscribes to a list, they can see the best posts from that list in their main news feed. Their lists also appear in the Interests section of their bookmarks. Being a social media pioneer Facebook has access to a lot of data of its user. But looking at the news feed in the same old Facebook wall proves to be monotonous and less attractive.
\par
The proposed software ''User Specific Dynamic E-Magazine'' is the answer to all the limitations of the products mentioned above. It removes all the problems with the above mentioned products also adding many other interesting and useful features. The solution is produced to the user in a very attractive GUI in the form of his personal e-magazine. This virtual magazine gives the feel of the actual real world magazine along with the interactive features. 

\section{Proposed Software Solution}
\label{b}
\begin{figure}[bpht!]
	\centering
		\includegraphics[height=14cm,width=12.5cm]{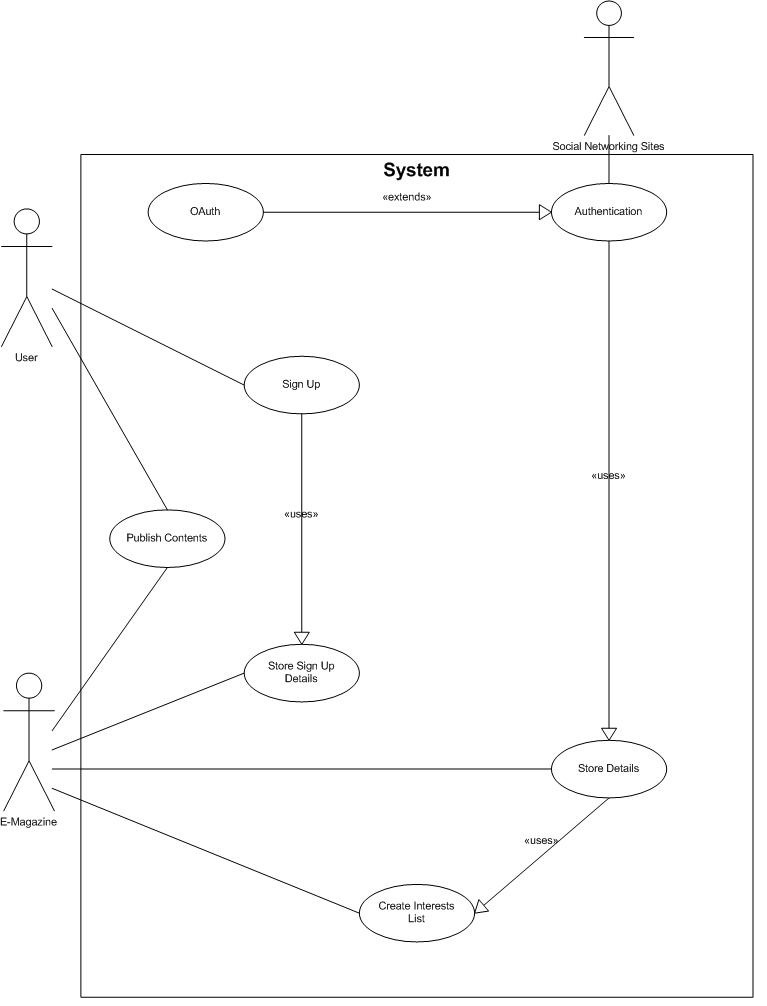}
	\caption{Use Case Diagram}
	\label{fig:f1}
\end{figure}

\begin{figure}[bpht!]
	\centering
		\includegraphics[height=14cm,width=12.5cm]{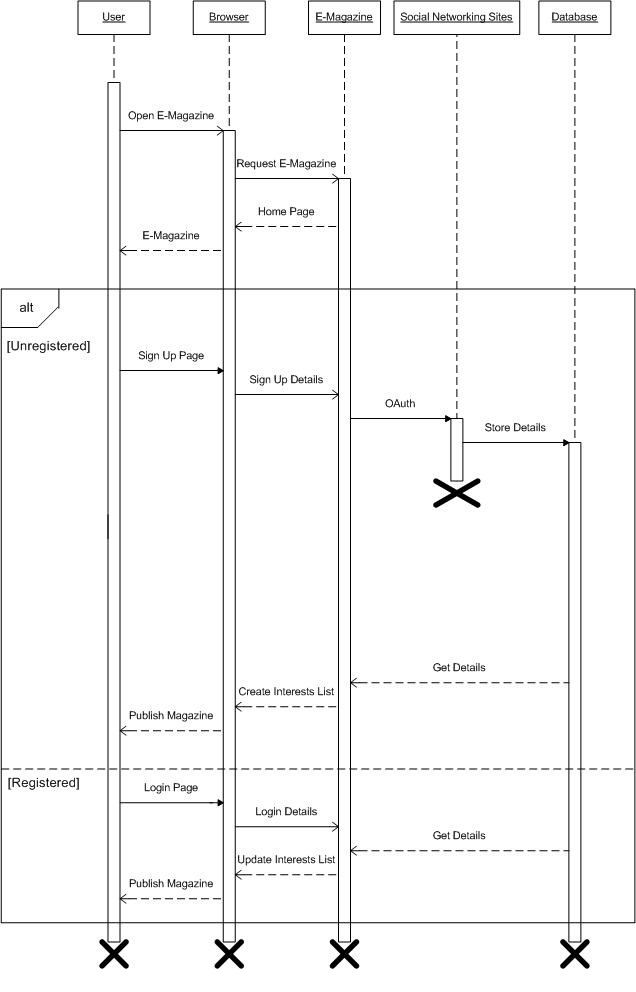}
	\caption{Sequence Diagram}
	\label{fig:f2}
\end{figure}

The purpose of this section is to present a detailed description of the design and development of the proposed solution ''User Specific Dynamic E-Magazine''. The overall process is described by the use-case and sequence diagrams shown in Fig.\ref{fig:f1} and Fig.\ref{fig:f2} respectively.

\subsection{Modules}
The proposed solution is divided into five modules which are listed below:
\begin{enumerate}

\item User data extraction by OAuth implementation.

\item	Data extraction in different media formats from the web.

\item User's data processing.

\item Creating Recommendation System.

\item Interactive GUI design.
\end{enumerate}

\subsubsection{User Data Extraction by OAuth Implementation}
The proposed solution integrates user's Facebook, Linkedin, Twitter and Google accounts with itself. Users share their data everyday on these social networking sites. This data can be videos, links posted or shared by the user, likes of the user, personal and professional information given by the user on these social networking sites. This private data of user is saved on the server of these social networking sites. To get access to these data, a user has to grant permission to the proposed application. This is done by a protocol called OAuth \cite{ath12}. OAuth implementation and integration details are given on the developers web-site for each of these social networking web-sites. Social networking sites provides the APIs to implement OAuth.
\par
OAuth is an open standard for authorization. It allows users to share their private resources (e.g. photos, videos, contact lists) stored on one site with another site without having to hand out their credentials, typically just by supplying username and password tokens instead. Each token grants access to a specific site (e.g., a video editing site) for specific resources (e.g., just videos from a specific album) and for a defined duration (e.g., the next 2 hours). This allows a user to grant a third party site access to their information stored with another service provider, without sharing their access permissions or the full extent of their data.
\par
So, when the users sign-up or register with ''User Specific Dynamic E-Magazine'' they are redirected to the social networking sites to grant permission. The proposed solution is registered as a web-application on each of these social networking sites. The social networking sites then provide the application with a consumer secret and consumer key. After the user grants the permission to the application, the social networking sites give them the access token. With the help of the access token the application can query the social networking sites to gain access to the user data.
\par
\begin{figure}[bpht!]
	\centering
		\includegraphics[height=14cm,width=12.5cm]{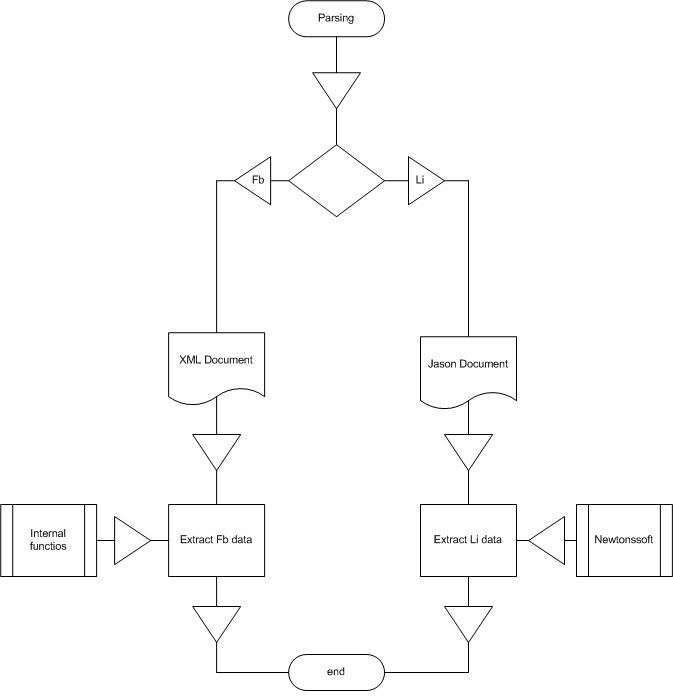}
	\caption{Parsing Facebook and LinkedIn Response}
	\label{fig:f3}
\end{figure}

The proposed software queries the database of these social networking sites using the access token by making a HTTP request and extracts the information. The response given by these social networking sites maybe in XML \cite{rfc1} or JSON \cite{rfc2} format. Figure \ref{fig:f3} depicts the process of data extraction from  Facebook and Linkedin.

\begin{figure}[bpht!]
	\centering
		\includegraphics[height=14cm, width=12.5cm]{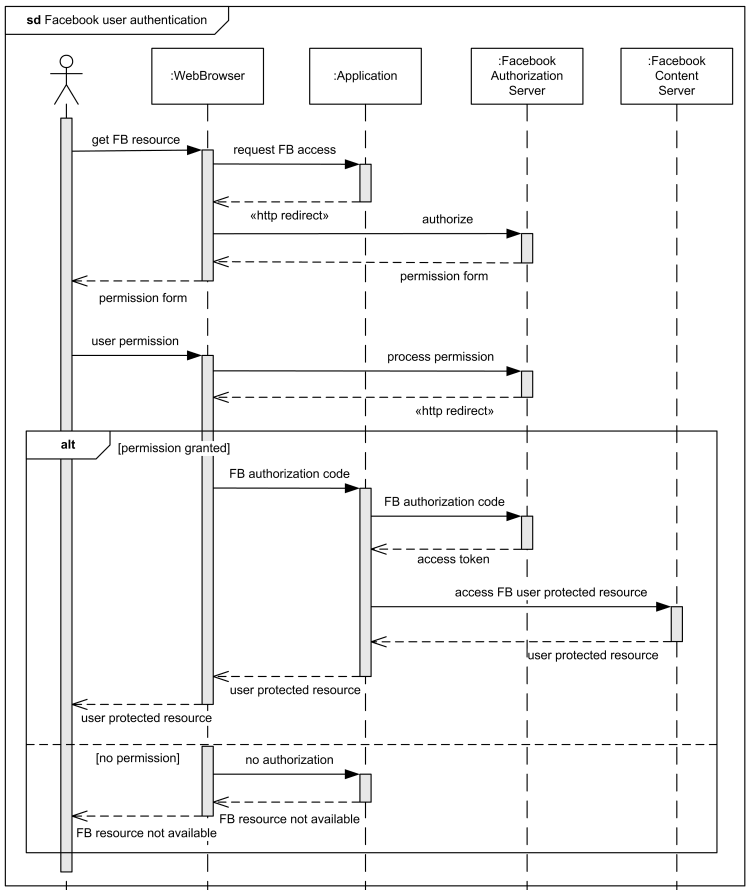}
	\caption{Facebook OAuth Protocol \cite{fbath}}
	\label{fig:f4}
\end{figure}

The application parses the data format and extracts the useful information of each user and stores it in its private database. The JSON returned by Facebook is parsed using an external library Newtonsoft \cite{new12}. The complexity of the OAuth standard can be seen and understood from the diagram shown in Fig.\ref{fig:f4}, which depicts the OAuth implementation for Facebook.

\subsubsection{Data Extraction in Different Media Formats from the Web}
\begin{figure}[bpht!]
	\centering
		\includegraphics[height=14cm,width=12.5cm]{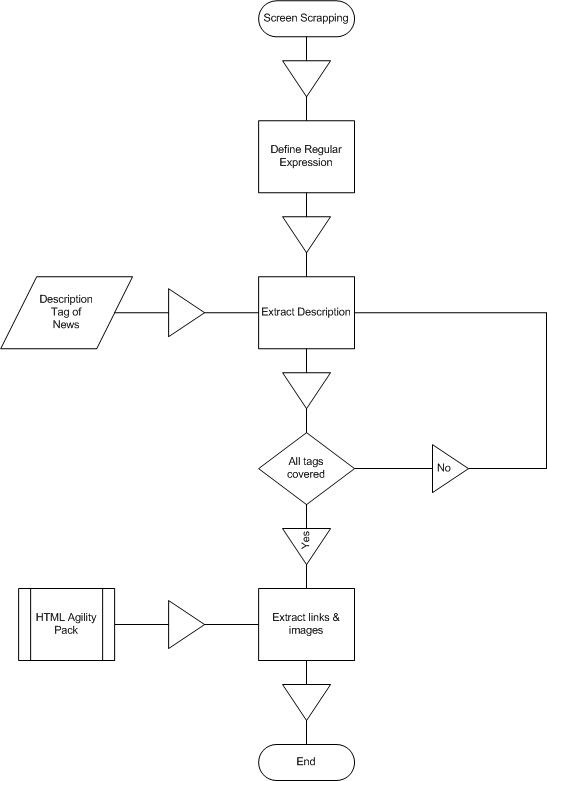}
	\caption{Screen Scrapping}
	\label{fig:f5}
\end{figure}

Data is picked up from more than 100 trusted sources, classified and stored in the application's database. Data pick-up is divided into two categories. Firstly data related to separate categories eg. (technology, entertainment, recreation, lifestyle etc.) are picked up from various sources and saved in the database after classification. The categories are divided into various sub-categories and each information is properly structured before storing into the database. 
The data is picked up from the web from various RSS feeds and then parsed to get the useful information. The parsing of data involves extracting the image tags, video tags, links, etc from the web-page, which is called Screen-Scrapping \cite{scr12}. In the proposed software Screen-Scrapping is done using Regular Expressions \cite{reg12} and an external library called HTML Agility Pack \cite{htmlag12}. Any new information appearing is evaluated for quality and accuracy, then classified and structured properly and then recorded in the database. Information can be of any media format, for example video, articles, news feeds, product reviews, images, etc. Only the links to the videos and images are saved in the database to save processing time and storage space. This process of classifying the data provides ease of access and ensures the accuracy of information given to each user. The process of screen-scrapping is depicted by the diagram shown in Fig.\ref{fig:f5}.

Secondly, the application is equipped with an in built search engine which makes use of HTTP request and response protocol to search the web for useful information based on a certain keyword, which defines the interest of any user. This type of data pick-up is required when the interests of the user is not found in the database. If the information or contents related to any user interest is not found in the database, then the web is crawled to search for the information. Once the information is found it is again evaluated for quality and then properly classified and saved in the database so that if the keywords crops up again the web is not searched or crawled.

\subsubsection{User Data Processing}
The data taken from the social networking sites are mapped into interests. For example, if a user has multiple likes related to sports, he will be definitely interested in sports, this can be further categorized to a particular sports, which can be indeed further categorized to interests related to any sports team or any sports personality. Similarly each data picked up from the social networking sites are mapped to user interests. These interests are again given some priority that is assigned weights according to the user's interest level. Interests level or weights keep on updating automatically according to the user's reading habits.
\par
The application calculates and updates user's interests and the corresponding interests level by analyzing the way the he interacts with the application itself. Users can also modify and update their interests and the corresponding interests level themselves. They can also define their own interests. Information and contents related to the interests are shown on the user's personal magazine in a concise form. User can click to navigate to the actual web-page from where the content had been taken if he is interested to know more or gain more information. This is one of the behaviors which is monitored and it helps to know the user better.
\par
The contents of the magazine keeps on updating automatically on the basis of time, relevance and interest level. The user can save any content in the applications database to view or read later. Any content published on user's personal magazine has the options to share it on any social networking sites like Facebook, Twitter, Linkedin, Google+ etc. Also the contents can be e-mailed by the user to any friend. User can rate the contents or post feedback or comment on the contents. These user behaviors can be used to track, calculate and update his interests.
\par
The application not only calculates the interests and makes the decision according to the things the user likes but also filters the contents which the user dislikes. This backward and forward process of filtering makes contents more and more accurate. The application learns from the user's reading behavior as explained above and gets smarter and more accurate.

\subsubsection{Creating Recommendation System}
The application recommends interest keywords and also the contents the user maybe interested in. Interests level are divided into three parts:
\begin{enumerate}

\item High Priority Interests. Contents related to these are published on the magazine.

\item	Mid Priority Interests. Contents are not published on these but recommended to the user instead, if the user wants he can assign them with High Priority.

\item Low Priority Interests. These keywords are just stored in the database waiting for the interest level to increase somehow. After a certain time if the interest level does not increase they are automatically flushed from the database. 

\end{enumerate}

There can be multiple keywords in the interests list of the user. The keywords with certain high interest levels are the only ones on which the contents are published at a particular time. When the user interacts with the application and his reading behavior is analyzed, it may lead to increase in the interest level of a particular keyword whose interest level is not yet high enough. For example the application calculated the interest of the user in ''Sachin Tendulkar''\footnotemark\footnotetext{Indian International Cricket Player}, but the interest level was not above the threshold to publish contents on this keyword, but if the user interacts with some contents published on his magazine related to the above keyword in some other way the interest level of the keyword is increased. Now the interest level is checked again and the keyword may fall under Mid Priority level and will be shown in the recommendations list.

\par
\begin{figure}[bpht!]
	\centering
		\includegraphics[scale=1]{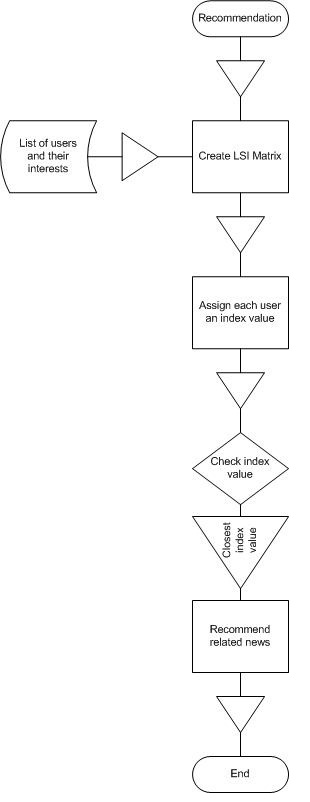}
	\caption{Recommendation System using LSI}
	\label{fig:f6}
\end{figure}

Users with similar interests are analyzed and similarities between them are calculated. If two users are found to be very similar the uncommon interests between them are recommended to the other user. This recommendation system is based on Latent Semantic Indexing (LSI) \cite{lsi12}. A LSI matrix of users and their corresponding keywords is formed. Then based on this matrix, a mathematical calculation is made and the matrix is decomposed using Singular Value Decomposition (SVD) to assign every user with a certain value called index. On comparing these index values we can understand which users are more similar and which users are dissimilar. Flowchart depicting the working of the LSI based recommendation system is shown in Fig.\ref{fig:f6}.

\subsubsection{Interactive GUI Design}
The GUI is built in the form of a magazine which has a page turning effect. It gives the feel of an actual magazine. It is built using HTML5 \cite{htm15}, CSS3 \cite{css23} and Jquery \cite{elec14}, which are the latest web development concepts.

\subsection{Pseudocode}
In this section we have given pseudocode of five algorithms which gives an overview of how news details are extracted or scrapped from any news website.
Algorithm \ref{alg1} shows how a request is sent and response is received in an XML form. This is then parsed to get the details.
Algorithm \ref{alg2} calls to separate modules to get particular details from the 'Description Tag' of the news RSS. Algorithm \ref{alg3} uses regular expressions to get the plain text from the description. Algorithm \ref{alg4} and algorithm \ref{alg5} are similar in nature, they use HTML Agility Pack to extract the links and images respectively.

\begin{algorithm} 
\caption{Screen Scrapping}
\begin{algorithmic}[1]
\STATE Select news RSS link
\STATE Send HTTP request
\STATE Receive HTTP response
\STATE Download RSS response as XML document
\STATE Load XML document
\FOR{$i=1$ to $end$}
\STATE \hspace{1.5cm} $Title$       = rssItems.Item($i$).SelectSingleNode("title")
\STATE \hspace{1.5cm} $Link$       = rssItems.Item($i$).SelectSingleNode("link")
\STATE \hspace{1.5cm} $Description$ = rssItems.Item($i$).SelectSingleNode("description")
\STATE \hspace{2.2cm} Get \textbf{DescriptionDetails($Description$)}
\STATE \hspace{1.5cm} $PubishDate$  = rssItems.Item($i$).SelectSingleNode("pubDate")
\ENDFOR
 	\end{algorithmic}
 			\label{alg1}
\end{algorithm}

\begin{algorithm}
\caption{DescriptionDetails($Description$)}
\begin{algorithmic}[1]
\STATE get DescriptionText($Description$)
\STATE get DescriptionLink($Description$)
\STATE get DescriptionImage($Description$)
\end{algorithmic}
\label{alg2}
\end{algorithm}

\begin{algorithm}
\caption{DescriptionText($Description$)}
\begin{algorithmic}[1]
\STATE Define regular expression pattern \textbf{($pattern1$)} to search text in Description
\STATE Define regular expression pattern \textbf{($pattern2$)} to search HTML tags in searched text
\STATE Search Description with $pattern1$
\IF{Search is True}
\STATE \hspace{1.5cm}Remove HTML tags from searched text using $pattern2$
\ENDIF
\RETURN $NewString$
\end{algorithmic}
\label{alg3}
\end{algorithm} 

\begin{algorithm}
\caption{DescriptionLink(Description)}
\begin{algorithmic}[1]
\STATE Define an array $anchor$ to store links
\STATE Load Description as HTML document $source$
\FORALL{anchor/link tags $i$ in $source$}
\STATE \hspace{1.5cm} $value$ = get $i$.AttributesValue["href"']
\STATE \hspace{1.5cm} add $value$ to $anchor$
\ENDFOR
\RETURN $anchor$
\end{algorithmic}
\label{alg4}
\end{algorithm} 

\begin{algorithm}
\caption{DescriptionImage(Description)}
\begin{algorithmic}[1]
\STATE Define an array $image$ to store images
\STATE Load Description as HTML document $source$
\FORALL{image tags $i$ in $source$}
\STATE \hspace{1.5cm} $value$ = get $i$.AttributesValue["src"]
\STATE \hspace{1.5cm} add $value$ to $image$
\ENDFOR
\RETURN $image$
\end{algorithmic}
\label{alg5}
\end{algorithm}

\subsection{Features}
This section discusses various features of the application.
\subsubsection{Rating the Contents}
Any information or contents published on the user's magazine can be rated by him from one to five. This can help the application to know the user better.
\subsubsection{Searching for Contents}
User also has an option to search for any contents he wants on the basis of the keyword. The contents can be of any media format. User can filter the search according to various parameters. The parameters can be specific media format eg. image, video etc., date of relevance of the content, from a particular or specific source. The search feature enables the user to expand his views and express his behavior to the application in a even better manner.
\subsubsection{Saving the Contents}
User can save any content he does not wants to lose or may read later. He can view his saved contents anytime he wants. The saved contents appear in the order of date saved. The contents can also be sorted on the basis of other attributes such as published date etc. There is also an option to filter the saved contents similar to the way searching contents can be filtered. This filtering saves users time when he has multiple items saved.
\subsubsection{Customization of Interests}
User can add or modify the list of his existing interests and their corresponding interest level. The top interests of the user appear in a very interactive GUI in the form of a Tag-Cloud \cite{elec19}. User can view the top interests calculated by the application for him on the Tag-Cloud and remove any existing interests from the list. Sliders on the GUI help the user to easily modify the interest level of any existing interests. Corresponding to the forward or backward motion of the sliders the size of a particular interest keyword on the Tag-Cloud increases or decreases. This gives a visual idea of the interest level of the user in each topic.
\par
This feature of interest customization enables the user to get any contents he wants and also helps the internal algorithms in the application to perform in a better, efficient and more accurate manner.

\subsubsection{Mailing the Contents}
User can mail the contents published on his magazine to any friend or even himself. 
\subsubsection{User Progress}
As soon as the user subscribes or registers, the application starts calculating his interests and starts learning from his behavior. With time the accuracy of the application increase as it learns the user behavior. This progress can be seen by the user as a progress bar in percentage. This helps the user to know his progress. If the progress is low, user can interact with the magazine and help it function even better.

\subsubsection{Sharing the contents on social networking sites}
Any content published on the user's magazine can be shared on his Facebook, Twitter, Linkedin or Google+ profile.

\subsubsection{Following Friends}
The interest list of any user is public by default, that is his Facebook friends present in this application can view his interest list. But the user has the option to make the list completely or partially private.
Users can see the interest list of their Facebook friends on the application. He can then follow the list completely or pick any keywords from the list and add to his own interest list. This can be done only if the interests list of the user is made public or even partially public.

\section{Conclusion}
\label{c}
Due to the ever increasing number of Internet users and published data on the web we are suffering from the problem of information explosion. Information explosion is also the reason for information pollution and information overload. This causes some serious consequences. So, a product was needed to overcome the problem of information explosion. In this paper we have developed a solution ''User Specific Dynamic E-Magazine'' which overcomes these problems and serves its users with qualitatively and quantitatively filtered information.
\par
The application ''User Specific Dynamic E-Magazine'' takes note of what the user is interested in and semantically filters its results according to those interests, even if they change. It does this using an intelligent semantic learning engine, which allows the application to continuously learn and adapts from the users interaction with it to become ever increasingly relevant. Each keyword or interest is semantically mapped to each information which may be in any media format. 
\par
User Specific Dynamic E-Magazine helps the people who are otherwise the victim of information overload. Because the application aggregates information around user's interests into distinct categories or single stream views, it removes the need to filter through masses of irrelevant information to find what user is actually looking for. Rather than looking for the information, the information comes to the user already aggregated and filtered in the form he wants it to be. The information given by the User Specific Dynamic E-Magazine is not only filtered but also quality filtered. This application is built using HTML5 and CSS3 so it is platform independent. Hence the application is fully compatible with Windows, IOS, and Android which enables it to be easily installed on tablets and smart phones. 


\bibliographystyle{IEEEtran}
\bibliography{myref}

\end{document}